\documentclass[conference]{IEEEtran}
\IEEEoverridecommandlockouts
\usepackage[colorinlistoftodos]{todonotes}
\usepackage{booktabs}
\usepackage{cite}
\usepackage{amsmath,amssymb,amsfonts}
\usepackage{algpseudocode}
\usepackage{graphicx}
\usepackage{textcomp}
\usepackage{xcolor}

\def\BibTeX{{\rm B\kern-.05em{\sc i\kern-.025em b}\kern-.08em
    T\kern-.1667em\lower.7ex\hbox{E}\kern-.125emX}}
\usepackage[linesnumbered,ruled]{algorithm2e}

\begin{document}

\title{Modelling of Rotor Wake using Viscous Vortex Particle Method \\

}

\author{\IEEEauthorblockN{Prithvi Dinesh Kewalramani}
\IEEEauthorblockA{\textit{Undegraduate } \\
\textit{Indian Institute of Technology Bombay}\\
 190010056@iitb.ac.in\\
 prithvik1969@gmail.com}
}

\maketitle

\begin{abstract}
This paper presents a comprehensive investigation into the modeling of rotor wake velocities, in a simplistic manner, using the Viscous Vortex Particle Method (VVPM). The study aims to accurately simulate wind velocities in the wake of helicopter rotors while comparing the VVPM with other established methods such as Momentum theory, Blade element theory, and the Free wake method. The VVPM is particularly promising due to its ability to simulate the interaction of rotor blade vorticities with a natural and efficient approach. The method employs a grid-free Lagrangian formulation, eliminating artificial numerical dissipation in the solution. To demonstrate the efficacy of the method, a Python code has been developed, enabling the visualisation of wake velocities for basic scenarios.
\end{abstract}


\section{Introduction}

A key challenge in helicopter design is modelling the wake velocities of the rotor, which can affect the helicopter's performance and stability. A rotor's wake is a complex phenomenon involving many factors, such as the rotor's geometry, speed, and the properties of the surrounding air. One of the critical reasons for understanding wake geometry is that the wake can affect the performance and stability of the helicopter \cite{10.1155/2019/9589415}. In addition, the wake can also affect other aircraft or objects in the surrounding environment. The wake generated by a helicopter can pose a risk to other nearby structures, particularly when they are flying at lower altitudes.

There are many methods to model the wake geometry and wind velocities, which range from straightforward to complicated, the accuracy of each varying with its complexity. Some methods discussed in this project include Momentum theory \cite{VANKUIK20151}, Blade element theory  \cite{BENINI2004957} , and the free wake method \cite{JAHS}, which can reasonably estimate rotor wake velocities and other parameters with somewhat less computational power. The more complex methods include Computational fluid dynamics \cite{https://doi.org/10.1002/we.458}, based on the Navier-Stokes equations, solving flow dynamics through numerical discretisation methods.

In the context of wake modeling, the Vortex Viscous Particle Method (VVPM) presents an improvement over the Free-vortex methods due to the latter's inherent limitations in capturing the full range of flow effects, due to their reliance on the potential flow assumption. In order to reach a more accurate solution, Free-vortex methods require empirical formulations, such as vortex decay factors or vortex core sizes, to approximate the rotor wake behaviour \cite{10.3390/app13031799}. Additionally, when significant viscous effects come into play, these effects are often either neglected, predetermined, or estimated by coupling the method with a boundary-layer model, thereby potentially introducing inaccuracies in the results. The majority of current computational fluid dynamics (CFD) methods, based on the Navier-Stokes equations, solve flow dynamics through numerical discretisation methods, such as finite difference. These methods tend to experience excessive dissipation owing to the numerical discretisation, leading to limitations in their accuracy while also requiring a tremendously high amount of computational power and time.

This project involves the development of a Python code implementation for the vortex particle method. The model adopts a grid-free Lagrangian formulation, devoid of any artificial numerical dissipation, allowing for greater flexibility in representing complex geometries and effectively handling unsteady flow phenomena. The VVPM code can accurately account for viscous effects and diffusion, resulting in a relatively accurate end result. 

The subsequent sections of this paper are as follows: Part II discusses other methods related to the viscous vortex particle method and lists the advantages and disadvantages of each. Part III provides an overview of the method, including the formulae and approach. Finally, we conclude the paper in Part IV and Part V, discussing the obtained results and providing a short conclusion to the experiment.

\section{Related Methods}

Simple methods such as Momentum theory, Blade element theory (BET) and Blade element momentum theory (BEMT) are a quick and easy way to get a ballpark figure for parameters of the rotor flow, such as the thrust and torque of the rotor, and the power required to drive it at the given conditions.

\subsection{Momentum Theory}
Momentum Theory  \cite{VANKUIK20151}  uses the principles of conservation of mass and momentum to determine the inflow and outflow velocities to the rotor. In the case of a rotor, momentum theory assumes that the rotor acts like a disc that draws air in and accelerates it, creating a wake behind the rotor. The theory only assumes uniform and axial inflow to the rotor using a control volume for the axial flow of air. It can roughly estimate the rotor’s performance, predicting the overall inflow velocity and the thrust and power required to drive the rotor. It is crucial to note that momentum theory provides a very simplified model, and does not accurately capture any complex phenomena occurring in the flow field around a rotor. It does not account for the number of blades, airfoil characteristics the blade shape or compressibility effects.

\subsection{Blade element theory and Blade element momentum theory}
Blade element theory \cite{BENINI2004957} is a simple method for calculating a rotor blade's aerodynamic forces and moments. It involves dividing the rotor blade into small segments and calculating each element's drag and lift forces. By adding up these values for each element on every blade of the rotor, we can determine the rotor's thrust, torque, and power values. One important advantage BET has over momentum theory is that it can account for various rotor parameters such as the number of blades, solidity, blade thickness and twist. The disadvantages of this method are that does not account for swirl losses and tip losses, and assumes uniform induced velocity. 

Hence we use a combination of BET and momentum theory instead, BEMT \cite{MAHMUDDIN20171123}, which addresses the uniform inflow assumption. BEMT equates the elemental thrust obtained from BET and the momentum theory to get a combined estimate for the inflow velocities at various locations. We can also combine Prandtl’s tip loss model to better represent the parameters that govern the functioning of the rotor. These methods work well with hover and forward, upward and downward flight, providing us fairly accurate results for thrust, power and torque. They are routinely used in more complex implementations of codes for calculating the above mentioned parameters, where the exact velocities at every point might be calculated by a method such as the Free wake model. Although it is important to note that these are steady state models, and are not versatile enough to be used in the case of accelerated flight.

\subsection{Vortex Methods}

Vortex methods are based on Prandtl’s lifting line theory, employing the Kutta-Juokowski theorem and
Biot Savart Law as its main governing equations, which provide us lift ($L$) and the induced velocity
($dV_{induced}$) as follows: 
$$ L=\rho_{\infty}V_{\infty}\Gamma$$ where $\rho_\infty$ and $V_\infty$ are the free stream density and velocity, $\Gamma$ is the circulation, and  $$dV_{induced}=\frac{\Gamma}{4\pi}\frac{d\vec{L}\times \vec{r}}{|r^3|}$$ where $\vec{r}$ is the distance vector from the vortex filament.
They provide a robust mathematical foundation for understanding the convergence characteristics of rotor flows, making them a valuable tool in predicting rotor performance. These models are computationally less expensive than grid-based methods like CFD, providing accurate solutions to wake velocity distributions. 

One significant advantage of vortex models is their ability to capture complex aerodynamic phenomena around a rotor, making them useful for predicting noise and vibration characteristics, as well as analyzing the effects of unsteady flow phenomena on rotor performance. These models are often used in combination with other methods, such as blade element theory or CFD, to provide a more comprehensive picture of the rotor's performance. While they have limitations, they can provide valuable insights into rotor system performance and are widely used in both research and industrial applications.

Comprehensive codes on the Free wake vortex method can account for tip losses, with the help of the Prandtl tip loss model as well as ground effect, using techniques such as the method of images, which involves creating a mirror image of the particles and their velocities and positions with respect to the ground plane. However, it does not come without its disadvantages. Most vortex methods have difficulty in modeling unsteady flows, since they assume that the flow is steady and uniform in the axial and tangential directions, which is often not be the case. Furthermore, it has limited ability to account for viscous effects, since the free vortex wake model is primarily a potential flow model. This limits its ability to accurately predict the behavior of the wake in regions where viscous effects are important, such as near the rotor blades or in regions of high turbulence.

\section{Proposed Method}

The viscous vortex particle method has been employed as a more promising and natural solution for simulating the interaction of rotor blade vorticity. This method utilizes a grid-free Lagrangian formulation, which avoids any artificial numerical dissipation in the solution process.

\subsection{Theory}
The main governing equation for VVPM is the Lagrangian description of the vorticity velocity form of the Navier Stokes equation:
\begin{equation}
    \frac{d\vec{\omega}}{dt}=\vec{\omega}.\nabla \vec{u} + \nu {\nabla}^2 \vec{\omega}
\end{equation}

where $\omega=\nabla \times u$ is the vorticity field associated with the velocity field $u$, $\nu$ is the kinematic viscosity , and $$\frac{d()}{dt}=\frac{\partial ()}{\partial t}+u.\nabla()$$ is the material derivative.

The vorticity field is represented by a set of N Lagrangian vector-valued particles as:
\begin{equation}
    \omega(\vec{x},t)=\sum^N_{i=1}\xi_\sigma(\vec{x}-\vec{x_i})\vec{\alpha_i}
\end{equation}
where $\alpha_i=\omega_iV_i$, with $V_i$ being the volume element size for each particle and $\omega_i$ being the vorticity of an individual element.
 $$\xi(\rho) = \frac{1}{ (2\pi)^{3/2}} e^{-\rho^2/2}$$ which is the Gaussian distribution function and $$\xi_\sigma(\vec{x})=\xi(|\vec{x}|)/\sigma^3$$$\sigma$ is a smoothing parameter which is the product of the particle overlapping parameter $c$ and the resolution parameter $h_{res}$. This parameter has been kept constant for the implementation of the code to reduce the working complexity.

For the convection of the vortex particles, we can define the equations for the velocity of the particles as follows:
\begin{equation}
    \vec{u}(\vec{x_i}, t)=-\sum^N_{j=1}\frac{1}{\sigma_{ij}^3}K(\rho)(\vec{x_i}-\vec{x_j})\times \vec{\alpha_j}
\end{equation}
with $\rho=|\vec{x_i}-\vec{x_j}|/\sigma_j$ as the non-dimensionalized distance parameter. $K(\rho)$ is the Biot-Savart Kernel, defined as $$K(\rho)=\frac{G(\rho)-\xi(\rho)}{\rho^2}$$
with $G(\rho)$ being Green’s function for the vector stream function, expressed as $$G(\rho)=\frac{1}{4\pi\rho}erf(\frac{\rho}{\sqrt{2}})$$
$erf$ is the error function, defined as $$erf(s)=2\int^s_0 e^{-\nu^2}\frac{d\nu}{\sqrt{\pi}}$$

Next, we look at the vortex stretching effect, which is controlled by the $\vec{\omega}.{\nabla} \vec{u} $ term in the governing equation. The vortex stretching effect is accounted for by using the so-called direct scheme, which directly multiplies the velocity gradient matrix with the particle vorticity as follows:
\begin{equation}
    \frac{d\vec{\alpha_i}}{dt}=\nabla\vec{u}(\vec{x_i},t).\vec{\alpha_i}
\end{equation}
where the velocity gradient can be obtained as:
$$\nabla\vec{u}(\vec{x_i}, t)= \sum^N_{j=1}\frac{1}{\sigma_{ij}^3}[\tilde{{\alpha_j}}][\nabla K(\rho)(\vec{x_i}-\vec{x_j})]$$
where $\tilde{\alpha_j}$ is the skew symmetric form of the vorticity vector, and the tensor for $[\nabla K(\rho)(\vec{x_i}-\vec{x_j})]$ is obtained as $$[\nabla K(\rho)(\vec{x_i}-\vec{x_j})]_{kl}=\delta_{kl}K(\rho) -\frac{1}{\sigma^2_{ij}} F(\rho)(\vec{x_{ik}}-\vec{x_{jk}})(\vec{x_{il}}-\vec{x_{jl}})$$ Here $$F(\rho)=\frac{[3K(\rho)-\xi(\rho)]}{\rho^2}$$

Now, to account for the viscous diffusion term in the equation,  $\nu {\nabla}^2 \vec{\omega}$, we get 
\begin{equation}
    \frac{d\vec{\alpha_i}}{dt}=\frac{2\nu}{\sigma^2_{ij}}\sum^N_{j=1}(V_i\vec{\alpha_j}-V_j\vec{\alpha_i}) \eta_{\sigma_{ij}} (\vec{x_i}-\vec{x_j})
\end{equation}
where the $\eta_{\sigma_{ij}}$ is taken as the Gaussian distribution function, same as the $\xi$ function above.

Hence with this, we get a complete picture of the velocities and vorticities of every particle in our simulation. (Please note that the derivation of the equations has not been mentioned here)

\subsection{Implementation on Python}

The above equations have been translated into our Python code as follows:

\begin{enumerate}
    \item Function Definitions: First, we establish the functions vital for the code's operation, namely the Green's function, Biot-Savart kernel, Gaussian distribution function, and other core elements.

    \item Initializing Particle Velocities: Subsequently, we define velocity values corresponding to initial particle positions. These velocities, in line with eqn. (3), stem from particle interactions and blade movement dynamics.

    \item Velocity Gradient Computation: We then employ eqn. (4) to determine velocity gradients, which are pivotal for subsequent computations.

    \item Combining Vorticity Gradients: Next, we integrate velocity gradients with vorticity gradients from viscous diffusion, as indicated by eqn. (5), updating the initial particle vorticity values. Using these values, we deduce the vorticity field equation as in eqn. (2). This yields the final vorticity values for discrete points, accounting for neighbouring particle influences.

    \item Output Generation: Finally, we obtain a tensor presenting computed velocity and vorticity values for each point.
\end{enumerate}

There are distinct code drivers for two scenarios: fixed-wing steady and level flight, and rotary-wing hover flight configurations. Presently, both implementations are rudimentary, assuming a simple linear wing alignment. This foundational approach provides a solid basis for future refinements.

\section{Results}

Here we discuss the outcomes yielded by the implementation of the VVPM in the contexts of both fixed-wing steady and level flight, and rotary-wing hover flight cases. This will serve as a foundation for discerning the wind flow dynamics in fixed-wing and rotary wing configurations, thus paving the way for a further comprehensive assessment of their respective aerodynamic characteristics.

\subsubsection{Fixed wing case}

For the fixed-wing steady and level flight case, the code driver utilizes a pre-defined grid of points on the x-y plane, arranged in an 11x10 matrix with each point spaced one unit apart from its neighboring point. The wing's span is aligned with the x-axis and is represented by a set of 11 equally spaced points. As the wing moves forward at a velocity of 0.1 units/second, the points are sequentially activated upon the wing's passage over them. This activation triggers their inclusion in the numerical calculations, allowing them to interact with other particles. Prior to activation, these points are merely stationary locations in space, and not yet considered actual particles.

This sequential activation scheme serves to simulate the shedding of vortex particles at regular intervals. The wing is discretised into bound circulation values, in the same pattern a wing in constant motion. The vorticity values are assigned to the wing points in a discrete manner, so as to obtain bound circulations on the wing in the following manner as illustrated in Fig. 1. The wing has constant vorticity as it moves forward, and it influences the other particles as it passes through.

\begin{figure}[h]
    \centering
    \includegraphics[width=0.7\linewidth]{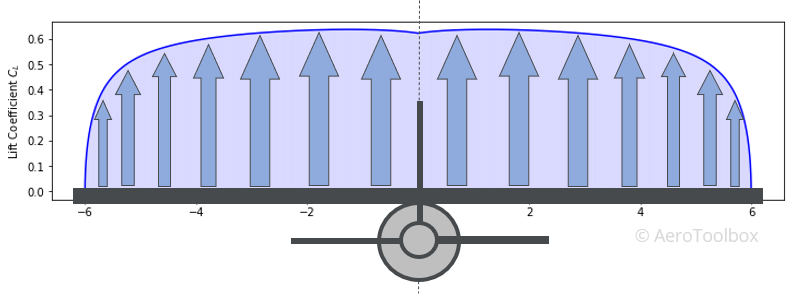}
   \caption{Bound circulation distribution on fixed wing}
\end{figure}

We get the following path for the trailing particles, as shown in Fig. 2 We do not see a satisfying trailing vortex, as one would expect, since there is no major viscous interaction, due to the very small number of particles. Though we do see the particles at the edges of the wings moving outwards.

\begin{figure}[h]
    \centering
    \includegraphics[width=0.7\linewidth]{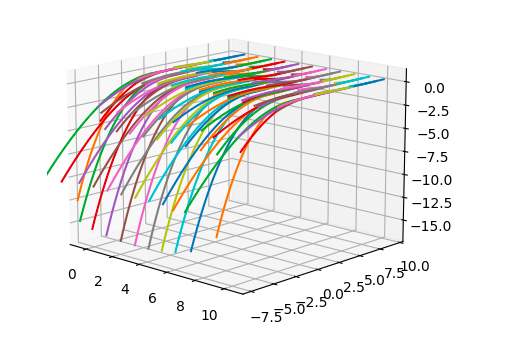}
   \caption{Pathlines for flow of particles in the fixed wing case}
\end{figure}

\subsubsection{Hover case}
In the case of rotary-wing hover , the driver code operates in a similar manner as that of the fixed wing case, with the added complexity of accounting for the wing's rotation. A set of particles is defined along 9 concentric circles, which establish the plane of motion for the rotor wing. Each point is spaced at an angle of 30$^o$ from its neighboring point. The wing is defined as a set of 9 points that rotate in a circular path, with the same activation mechanism for each point in the path as in the fixed wing case. The challenge arises from the constant angular velocity of the wing, necessitating a constant relative motion of the points with respect to the wing. To achieve this, the vorticities of both the points and the wing are updated at each time step by the angle moved by the points and the wing, respectively.

The wing's vorticities are maintained constant throughout the simulation and are discretized to obtain a circulation distribution according to the shape shown in Fig. 3.

\begin{figure}[h]
    \centering
    \includegraphics[width=0.7\linewidth]{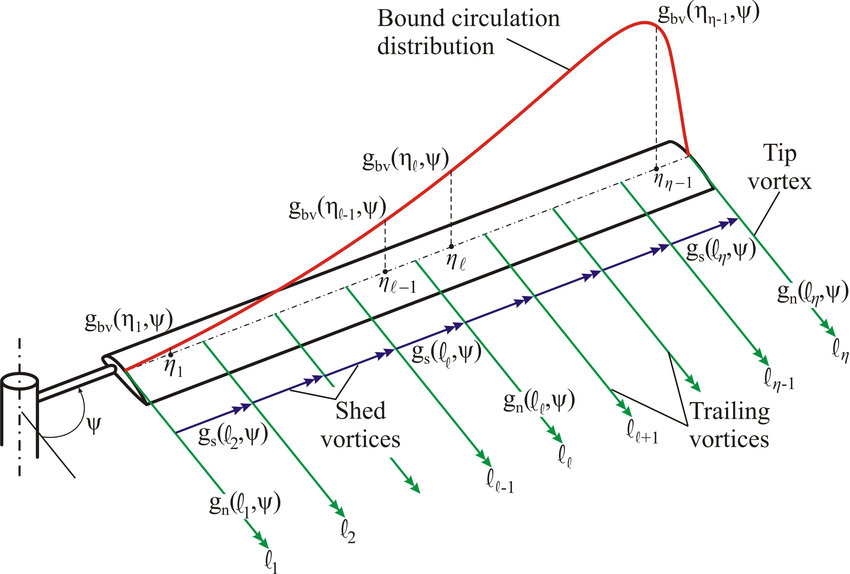}
   \caption{Bound circulation distribution on rotor wing}
\end{figure}

The simulation is run for a single rotation in the code to examine the outcomes. For the sake of clarity, only 5 of the 9 concentric circles are depicted in the diagram. 
The obtained results demonstrate a congruence with practical simulations, lending support to the methodology. However, it is worth noting that a small number of particles counter to the anticipated downward trajectory. This could potentially be attributed to viscous interactions, thereby inducing marginal deviations from their anticipated trajectories.

\begin{figure}[h]
    \centering
    \includegraphics[width=0.7\linewidth]{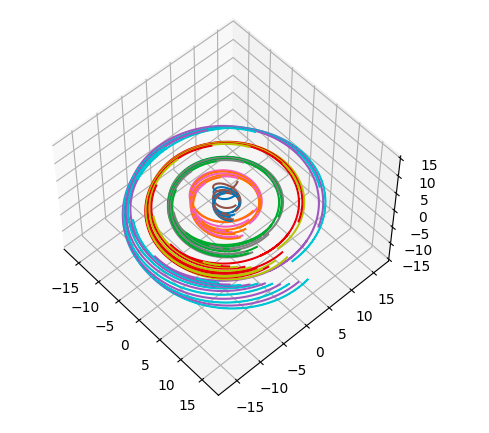}
   \caption{Pathlines for flow of particles in the hover wing case (top view)}
\end{figure}

\begin{figure}[h]
    \centering
    \includegraphics[width=0.7\linewidth]{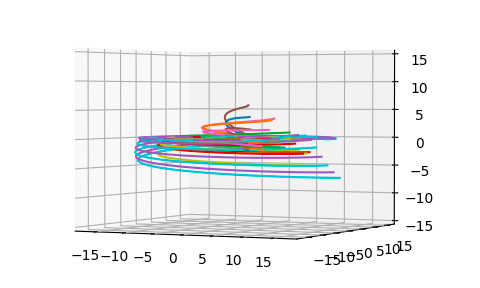}
   \caption{Pathlines for flow of particles in the hover wing case (side view)}
\end{figure}

\section{Concluding remarks}

With this research, we have successfully implemented the Viscous Vortex Particle Method (VVPM) within a Python framework to analyze the vortex wake patterns for rotary wing movement. By translating theoretical equations into practical code, we gained valuable insights into the behaviors of these effects. 

The analysis of the fixed-wing case revealed the basic workings of particle activation and their interactions, shedding light on the interplay of vorticity and motion.  Though the analysis for rotary-wings was elementary, with no added adjustments, results for the hover case closely aligned with anticipated outcomes, giving weight to the vortex particle method theory. In summation, the successful implementation of the VVPM underscores its potential as a powerful tool for studying complex aerodynamic phenomena, with the outcomes paving the way for better explorations in this domain.

Further iterations of the code will include added functionality such as tweaking wing parameters, variation of rotor velocity, as well as forward and upward flight. In addition, the use of techniques such as the tree-code method to reduce time complexity could be added in the future. Overall the code addresses a few major drawbacks to the free-vortex method and gives a more realistic outlook to the particle interactions, by incorporating viscosity and diffusion effects.

\section{Acknowledgements}

I would like to express my gratitude to Prof. Dhwanil Shukla at IIT Bombay, whose guidance and support have been instrumental in the completion of this project. His vast knowledge and expertise in rotor aerodynamics provided invaluable assistance whenever I faced challenges during the course of this work. Additionally, his introduction to more advanced concepts proved to be a great learning opportunity. I am also grateful for his efforts in helping me understand the papers on which this project is based. Without his encouragement and constant support, this project would not have been possible. Lastly, I would like to thank him for taking the time to review my code and providing feedback to help me correct errors and improve the quality of my work.

I would also like to thank Sharookh Ali MM for providing me with the code for this project. His explanations were instrumental in shaping my ideas and improving the scope of my work.

\bibliographystyle{apalike}
\bibliography{main}

\begin{thebibliography}{}

\bibitem[Bagai and Leishman, 1995]{JAHS}
Bagai, A. and Leishman, J.~G. (1995).
\newblock Rotor free‐wake modeling using a pseudo‐implicit technique —
  including comparisons with experimental data.
\newblock {\em Journal of the American Helicopter Society}, 40, Number
  3:29--41.

\bibitem[Benini, 2004]{BENINI2004957}
Benini, E. (2004).
\newblock Significance of blade element theory in performance prediction of
  marine propellers.
\newblock {\em Ocean Engineering}, 31(8):957--974.

\bibitem[Mahmuddin, 2017]{MAHMUDDIN20171123}
Mahmuddin, F. (2017).
\newblock Rotor blade performance analysis with blade element momentum theory.
\newblock {\em Energy Procedia}, 105:1123--1129.
\newblock 8th International Conference on Applied Energy, ICAE2016, 8-11
  October 2016, Beijing, China.

\bibitem[Park et~al., 2019]{10.1155/2019/9589415}
Park, J., Kim, D., Chae, S., Lee, Y., and Go, J. (2019).
\newblock Vibration and performance analyses using individual blade pitch
  controls for lift-offset rotors.
\newblock {\em International Journal of Aerospace Engineering}, 2019:1--13.

\bibitem[Salinas et~al., 2023]{10.3390/app13031799}
Salinas, M.~F., Botez, R.~M., and Gauthier, G. (2023).
\newblock New validation methodology of an adaptive wing for uav s45 for fuel
  reduction and climate improvement.
\newblock {\em Applied Sciences}, 13:1799.

\bibitem[Sanderse et~al., 2011]{https://doi.org/10.1002/we.458}
Sanderse, B., van~der Pijl, S., and Koren, B. (2011).
\newblock Review of computational fluid dynamics for wind turbine wake
  aerodynamics.
\newblock {\em Wind Energy}, 14(7):799--819.

\bibitem[{van Kuik} et~al., 2015]{VANKUIK20151}
{van Kuik}, G., Sørensen, J., and Okulov, V. (2015).
\newblock Rotor theories by professor joukowsky: Momentum theories.
\newblock {\em Progress in Aerospace Sciences}, 73:1--18.
\newblock “Special N.E. Joukowsky Volume” “This special volume contains
  two comprehensive papers on the history of rotor aerodynamics with special
  emphasis on the original pioneering contributions of Professor N.E.
  Joukowsky.”.

\end{thebibliography}

\end{document}